\documentclass[a4paper,fleqn,usenatbib]{mnras}
\usepackage[T1]{fontenc}
\usepackage{ae,aecompl}
\usepackage{graphicx}
\usepackage{rotating} 
\usepackage{epsfig} 
\usepackage{amsmath}
\usepackage{amssymb}
\usepackage{euscript}
\usepackage{pdflscape}
\usepackage{url}

\title[Spectrum of the accretion shock]{The structure and spectrum of the accretion shock in the atmospheres of young stars}

\author[A.~Dodin]{Alexandr~Dodin \thanks{E-mail: dodin\_nv@mail.ru} 
\\
Sternberg Astronomical Institute, Moscow M.V. Lomonosov State University,
Universitetskij pr., 13,  Moscow, 119992, Russia}

\date{Accepted ~~~~~~~~~~~~~~~~~~~~~~~~~~~~~~~~. Received ~~~~~~~~~~~~~~~~~~~~~~~~~~~~~~; in original form }
\pubyear{2018}

\begin{document}
\label{firstpage}
\pagerange{\pageref{firstpage}--\pageref{lastpage}}
\maketitle

\begin{abstract}

The structure and spectrum of the accretion shock have been self-consistently simulated  for a wide range of parameters typical for Classical T Tauri Stars (CTTS). Radiative cooling of the shocked gas was calculated, taking into account the self-absorption and non-equilibrium (time-dependent) effects in the level populations. These effects modify the standard cooling curve for an optically thin plasma in coronal equilibrium, however the shape of high-temperature ($T>3\times10^5$\,K) part of the curve remains unchanged. The applied methods allow us to smoothly describe the transition from the cooling flow to the hydrostatic stellar atmosphere.  Thanks to this approach, it has been found that the narrow component of \ion{He}{ii} lines is formed predominantly in the irradiated stationary atmosphere (hotspot), i.e. at velocities of the settling gas $<2$\,km\,s$^{-1}$. The structure of the pre-shock region is calculated simultaneously with the heated atmosphere. The simulation shows that the pre-shock gas produces a noticeable emission component in \ion{He}{ii} lines and practically does not manifest itself in \ion{He}{i} lines ($\lambda\lambda$ 5876, 10830\,{\AA}). The UV spectrum of the hotspot is distorted by the pre-shock gas, namely numerous red-shifted emission and absorption lines overlap each other forming a pseudo-continuum. The spectrum of the accretion region at high pre-shock densities $\sim10^{14}$\,cm$^{-3}$ is fully formed in the in-falling gas and can be qualitatively described as a spectrum of a star with an effective temperature derived from the Stefan-Boltzmann law via the full energy flux.
\end{abstract}


\section{Introduction}

Classical T Tauri stars (CTTS) are young low mass stars, accreting matter from a protoplanetary disc. According to the current paradigm (see e.g. the review by \citealt*{Hartmann16}), the observational properties of these stars can be explained in the frame of a magnetospheric accretion scenario \citep{Koenigl91,Romanova03}. In this model,
the accretion flow at distances smaller than a few stellar radii is governed by a strong stellar magnetic field. Falling down toward the star along the field lines, the plasma is accelerated up to the free-fall velocity at the stellar surface. The kinetic energy of the flow is converted into heat in the strong shock and escapes in the form of radiation. 

The radiation from the shock cooling region, concentrated mainly in the UV and X-ray regions, irradiates the stellar atmosphere on one side and the free-falling gas on the other side of the shock. 
\citet{CG98} and \citet{Dodin12} have shown that the heated atmosphere or, in other words, the hotspot is a source of continuous emission and numerous emission lines, which reduce the depths of absorption lines in the spectra of CTTS compared to the spectra of main-sequence stars of the same spectral types. Matching the observed excess emission in the visible and UV regions with a suitable model is the main method used for the determination of the accretion rates \citep{Ingleby13}. Note that the accretion rate is measured only on visible part of the stellar surface and can differ from the total accretion rate, but if the rotational axis of the star allows observation of most of the surface, then the total accretion rate can be retrieved from spectra obtained at various rotational phases. Certainly, this requires a stationary picture of accretion, which is probably not always realized \citep*{Blinova16}. Nevertheless, there are stars where the spectral accretion signatures are stable on time-scales of a few rotational periods; one example is V2129 Oph \citep{Alencar12}. 
The stable accretion and dense phase coverage allow one to not only deduce the rotational modulation of the accretion rate, but also obtain a map of accretion parameters on the stellar surface by using the Doppler imaging technique \citep*{Piskunov90}. To apply this technique, it is necessary to know an angular distribution of specific intensity $I_{\nu}(\mu)$ for each line involved in the analysis. The efficiency of this approach has already been demonstrated in the magnetic mapping of young stars (see e.g. \citealt{Donati07,Donati08,Donati11,Donati}). However, it should be noted that magnetic imaging has been carried out without proper account for a hotspot, which can distort the obtained results. The specific intensity $I_{\nu}(\mu)$ for the hotspot has been recently calculated by \citet{Dodin12} and \citet{Dodin15}; the current work is in part a continuation that more accurately treats the radiation, which heats the hotspot, and includes self-consistently the pre-shock, which probably begins to dominate in the spectrum at the pre-shock densities $N_0\sim10^{13}$\,cm$^{-3}$ \citep{Lamzin95}. 
But even at moderate densities, the in-falling gas manifests itself in spectra as a red-shifted component of some strong lines (e.g. \ion{C}{iv} $\lambda$1550\,{\AA}, \ion{He}{ii} $\lambda$1640\,{\AA} \citep{Ardila}, \ion{He}{i} $\lambda$5876\,{\AA}, \ion{O}{i} $\lambda$7773\,{\AA} \citep{Petrov}, and other lines).
Large Doppler shifts of these components make them informative for a diagnostic of the accretion geometry;
however, to make such a diagnostic, a quantitative theory of formation of these components should be constructed.
The above serves as a motivation for a detailed calculation of the structure and spectrum of the pre-shock region.

There are several works dedicated to the modelling of spectral lines in the free-falling gas.
\citet*{Muzerolle01}, \citet{KR13} have considered the formation of \ion{H}{i} and \ion{Na}{i} lines in the accretion column 
far from the shock front. The important limitation of these models is an uncertainty in the calculation of the thermal structure, related to the unknown processes responsible for the gas heating. Whatever the processes were, the temperatures considered in these models 
$<10\,000$\,K  allow to reproduce the observed inverse P-Cygni profiles of \ion{H}{i} and \ion{Na}{i}, but cannot produce \lq{hot lines}\rq{} such as \ion{He}{i-ii}, \ion{C}{iv}. In the case of \ion{He}{i}, this problem can in principle be solved if we include in the calculation coronal X-rays, even small amounts of which can lead to appearance of \ion{He}{i} lines \citep*{KR11}. 

It is natural to suppose that the hot lines should be formed in the pre-shock region irradiated by the shock emission. Line formation in the pre-shock region due to photoionization has been considered by \citet{Lamzin03}, who has calculated the specific intensity $I_{\nu}(\mu)$ for some emission lines.
In our work, we present a new model of the shock region, which takes into account the line and continuum emission originating from the pre-shock and post-shock regions as well as the hotspot.
The in-falling gas is irradiated by the same radiation as the heated atmosphere; moreover, both regions have similar physical parameters and as a consequence, the emergent spectra of these regions are similar, i.e. they can absorb and re-emit radiation of each other. To take into account this radiative interaction, both regions will be calculated simultaneously.

The ionizing radiation of the post-shock has already been calculated by \citet*{Lamzin98} and has been used in previous work dedicated to the modelling of the hotspot \citep*{Dodin15}. However, this radiation is given in the form of flux $F_{\nu}$, while we need a specific intensity $I_{\nu}(\mu)$ (see discussion in \citealt{Dodin15}). The angular distribution of $I_{\nu}(\mu)$ is not universal for the whole spectrum, but depends on the optical thickness at each frequency. To obtain the correct angular distribution, a new calculation of the post-shock is needed. 

The necessity of such calculations is also caused by the fact that the post-shock smoothly passes in the heated atmosphere, while in previous studies (e.g. \citealt{Lamzin98,Dodin15}) these regions were simulated separately by substantially different methods. An ambiguity of the division into these regions leads to uncertainties in the line profiles formed on their boundary, such as \ion{C}{iv}, \ion{He}{ii}, etc. In the other words, works dedicated to the modelling of the post-shock spectra do not take into account the contribution of the heated atmosphere and vice versa. Here we consider a model in which the cooling flow and the stellar atmosphere are calculated in a single approach that allows us to reach a smooth transition from one to another.

To calculate the ionizing radiation, we will suppose that the shock is stationary. On the one hand, the theory predicts  \citep{Koldoba, Sacco} that the shock must be unstable in the case of the standard cooling curve for an optically thin plasma; on the other hand, recent observational data of young stars do not show any evidence of such an instability \citep{Drake, Gunther}. The absence of observed manifestations of the instability can be explained if the shock is not uniform. Due to the strong magnetic field, the accretion flow can be represented as a set of flux tubes, in each of which the shock front oscillates independently of the others. Moreover, the frequencies of these oscillations may differ if the gas densities in these tubes differ \citep{Matsakos}. An alternative explanation for the absence of the instability is the difference between the real cooling curve and the standard one. The most obvious causes of such deviations are non-equilibrium (time-dependent) effects in the level populations \citep*{SD93, Lamzin98} and self-absorption of the produced radiation. In the present work, we will consider how these effects transform the cooling curve.

\section{Modelling the structure of the post-shock region}

The steady-state cooling flow is modelled by solving HD
equations of conservation of mass, momentum, and energy: 
%
\begin{equation}\label{mass}
       \rho V = \rho_0 V_0 \equiv u_0,   
\end{equation}       
%
%
\begin{equation}\label{mom}
       P+\rho V^2 = \rho_0 V_0^2 + P_0, 
\end{equation} 
%
%
\begin{equation}\label{en}
      u_0 \frac{ {\rm d} E}{ {\rm d} m}= H - C. 
\end{equation}       
%
Gravity and heat conductivity are neglected. Here $\rho$ is the mass density. Alternatively, we will use a number density~$N,$ which is related to $\rho$ as: $\rho=2.11\times10^{-24}$(g)$\times N,$ for the solar elemental abundances. 
$V$ is the plasma velocity;
$P$ is the thermal pressure; and
$m$ is the mass coordinate.
We denote the pre-shock values by subscript $0.$
$E$ is the specific energy, which for an ideal monatomic gas can be expressed as
%
\begin{equation}\label{endef}
     E =  \frac{V^2}{2} + \frac{5}{2}\frac{P}{\rho}+\varepsilon, 
\end{equation}     
%
\begin{equation}\label{epsdef}
    \varepsilon = \frac{1}{\rho}\sum\limits_l {N_l E_l},
\end{equation} 
%
where $N_l$ are the populations of all atomic levels of all elements in all ionization stages, and
$E_l$ are the energies of the levels. The variables
$H = 4\pi \int{\chi_{\nu}J_{\nu}d \nu},$
$C = 4\pi \int{\chi_{\nu}S_{\nu}d \nu}$ are the heating and cooling rates of the gas per unit mass.
$\chi_{\nu}$, $J_{\nu}$, $S_{\nu}$ denote correspondingly the opacity, the mean intensity, and the source function at the frequency $\nu.$

The radiation field is determined by the radiative transfer equation (RTE), which is solved by the short characteristic method \citep*{OK87}. The spectra are calculated from 1\,{\AA} to 230\,$\umu$m on an adaptive grid with $\sim 10^{5}$ frequency points. The Thomson optical thickness for all models is $<10^{-4},$ therefore the scattering on free electrons is negligible and ignored.

Because in the cooling flow the statistical equilibrium (SE) may be absent, to calculate the fractional level populations $n_i$ the corresponding system of ODEs for each chemical element should be solved
%

\begin{multline}\label{se}
     u_0 \frac{ {\rm d}  n_i}{ {\rm d} m} = 
     \sum\limits_{j=1}^{K} (R_{ji}+C_{ji})n_j - n_i \sum\limits_{j=1}^{K} (R_{ij}+C_{ij}), \\
      \shoveright{ i=1,K-1,  
       \qquad \sum\limits_{i=1}^K n_i = 1,}
\end{multline} 
%
where $K$ is the number of the last considered level, i.e. the ground state of the last considered ionization stage of the element. $R_{ij}$ and $C_{ij}$ are the radiative and collisional rate coefficients from level i to level j.
After $n_K$ is expressed from the last equation via the rest of the $n_i,$ the system of stiff differential equations are solved by using a variable-coefficient ordinary differential equation solver \citep*{Brown}. Recall that the populations in SE are defined by linear equations, which are obtained from Eq.~(\ref{se}) by setting $u_0 \frac{ {\rm d} n_i}{ {\rm d} m}=0.$ Finally, the level populations
are calculated as follows
$N_i = \xi_{\rm el} N n_i,$
where
$\xi_{\rm el}$ is the abundance of the element $\rm el$;
$N$ is the number density of all atoms and ions.
The self-consistent solution of the RTE and equations (\ref{se}) is achieved by the accelerated $\Lambda$-iteration method.

To calculate the temperature distribution $T(m)$ in the post-shock region that satisfies equations (\ref{mass}-\ref{en}), the modified $\Lambda$-correction \citep{Dodin17} can be applied. However, simulations with this correction scheme show that temperature in the region of interest drops monotonically, which allows the use of a more simple method. Satisfaction of equation (\ref{en}) is achieved by calculation of the mass coordinates $m(T),$ which correspond to the predefined temperature grid evenly spaced on a logarithmic scale from $T_{\rm max}$ to $T_{\rm min}$. The coordinates $m$ are calculated from the equation (\ref{en}), which can be rewritten as
%
\begin{equation}\label{meq}
     {\rm d} m=\frac{u_0}{ H-C }  {\rm d} E, 
\end{equation}     
%
where the right side is known from the previous iteration. The convergence is usually reached after five iterations. 
$T_{\rm max}$ is defined from the initial conditions at the shock front.
$T_{\rm min}$ is determined by the condition that the gas reaches the radiative equilibrium ($H\approx C$).
As we will see later, the gas, cooled down to $T_{\rm min}$, loses  almost all its velocity ($V \lesssim 1$\,km\,s$^{-1}$) and reaches not only the radiative equilibrium but also the statistical  equilibrium, i.e. below this temperature the gas can indeed be considered as a part of the hotspot (see the next section).

Generally speaking, just behind the shock front the electron and ion temperatures are different. But in the regions where this difference is significant, the gas emits a negligible ($\lesssim 10^{-3}$) part of its energy \citep{Lamzin98}. We do not consider such thin structures at the shock front, which allows us to simplify the simulation.

\subsection{The initial conditions}

The post-shock cooling zone is linked to the models of the pre-shock gas and hotspot by the initial conditions for the equations (\ref{se}, \ref{meq}) and the RTE.

Obviously, the level populations just before the shock should be taken as initial conditions for Eq. (\ref{se}).
However, it turned out that just behind the shock, the populations reach equilibrium values at short distances due to high collisional ionization rates. The distance where this equilibration occurs nearly corresponds to the distance where the electron and ion temperatures become equal. This coincidence is not accidental, because both scales are determined by the time of electron-ion collisions. Because for our simulation this region can be ignored, we set the SE populations as initial conditions for Eq. (\ref{se}).

The initial values of the variables $V, P, \rho$ are determined by the jump conditions (see Appendix \ref{A1}).
The initial velocity is
%
\begin{equation}\label{vin}
V_{\rm in} \approx 
\frac{V_0}{4}\left( 1+ 5\frac{P_0}{\rho_0V_0^2} - \frac{8}{3} \frac{\varepsilon_{\rm in}-\varepsilon_0}{V_0^2} \right),
\end{equation}
%
where the last correction is due to the fact that when the gas reaches the first grid node, some part of its thermal energy has been spent on ionization and excitation of atoms and ions. The pre-shock values $P_0$ and $\varepsilon_0$ at the first calculation are set to zero and become known when the pre-shock model is calculated. $P_{\rm in}$ and $\rho_{\rm in}$ are easily calculated from Eqs. (\ref{mass}, \ref{mom}), and the initial temperature $T_{\rm max}$ is deduced from the ideal gas equation $P=(N+N_e)kT$.
The initial condition for the energy equation (\ref{meq}) can be obtained from Eq. (\ref{endef}) for the pre-shock parameters.

The initial condition for the RTE at the shock front is the intensity emerging from the pre-shock toward the star, and for the opposite boundary of the cooling flow it is the intensity emerging from the heated atmosphere. Because the hot gas is mostly transparent for the radiation from the hotspot, these conditions have almost no effect on the structure of the shock wave but allow a more smooth transition for the level populations from the cooling flow to the hotspot to be obtained. 
At the first step, while the models of the free-falling gas and hotspot are not constructed, we assume that these regions do not emit and have a zero temperature that allows calculation of the structure and spectrum of the cooling flow, and then calculation of the structure and spectrum of the pre-shock gas and hotspot, and a recalculation of the cooling flow with new initial conditions. To reach convergence and a smooth transition at the conventional boundary between the cooling flow and the hotspot, it is sufficient to perform 2--3 such global iterations.
All three regions can in principle be calculated together, however, we isolate the calculation of the cooling zone in a separate unit, which is linked with surroundings through the boundary conditions. This complication is necessary due to different convergence rates of the post-shock model and pre-shock--hotspot model. The calculation of the post-shock model takes about 10 times fewer $\Lambda$-iterations than the pre-shock--hotspot model, but each such iteration is about 100 times longer. Thus, the separation of the code into two units greatly reduces CPU time.

\section{ Modelling the structure of the pre-shock region and the heated atmosphere}

Using well-developed methods of the theory of stellar atmospheres, \citet{Dodin15} has calculated the non-LTE structure and spectrum of the hotspot, taking the pre-shock gas into account in a simplified manner. In this work we perform a simultaneous calculation of both regions by using the same methods as in \citet{Dodin15}. 
Referring to this paper for details, we recall the main assumptions used for the modelling. The geometry is plane-parallel. The thermal structure is determined by the radiative equilibrium. The level populations of H, He, C, and O are in SE, while the rest of the elements (from Li to Zn) are in LTE.
In this paper, we will mainly concentrate on the spectra of hydrogen and helium, whereas the departures from LTE for C and O will be considered only for accounting for their impact on the H and He lines. The RTE is solved including line and continuum opacity sources, scattering is neglected.

All equations and methods are the same for both regions except for the equation of hydrostatic equilibrium, which is solved for the hotspot only, while the pre-shock is considered as a slab of gas with constant density and velocity. An optically and geometrically thin source of ionizing radiation is placed between the pre-shock and hotspot.
The intensity of ionizing radiation $I^{sh}_{\nu}(\mu)$ is calculated by the methods described in the previous section. The shock radiation ionizes and heats the surroundings, which also begin to radiate, predominantly in the UV and visible spectral regions. In order to perform the simultaneous calculation of the pre-shock and the hotspot, the numerical solution scheme of the RTE is modified in such a way that the intensity of the ionizing radiation $ I^{sh}_{\nu}(\mu)$ is added to the hotspot radiation (for $\mu>0$) or to the pre-shock radiation ($\mu<0$) at a depth located between 
the last grid point of the pre-shock and the first point of the hotspot.

\subsection{The Doppler shifts}

Because two regions move towards each other, the radiation of each, observed from the opposite, is blue-shifted by $\Delta \nu = \nu V_0 \mu /c$, where $\mu>0$ is the cosine of the angle between the direction of light propagation and the velocity vector, which is parallel to the normal, therefore this $\mu$ coincides with $|\mu|$ in the RTE.

As in \citet{Dodin15}, to take into account the line blanketing, we use the opacity sampling technique, which consists of choosing a fixed grid of $10^4$ - $10^5$ frequencies. This grid is too coarse for the calculation of the spectrum, but is sufficient for accounting for the line opacity. In an addition to this grid, the lines of elements, treated in non-LTE, are calculated on a special grid for each line. This method can be easily applied to our case of moving gas, if we define a dynamic grid for $\mu$ for each frequency $\nu_i$ by the relation
$\mu_k=(\nu_{i+k}-\nu_i) c/V_0$
in which $k$ starts from 0 and the last $k$ is determined by the condition $\mu_k \leq 1.$ 
Even for the usual number of frequency points ($> 30\,000 $) the angular grid is sufficiently dense to compute various moments of the radiation field. The RTE along each $\mu$ is solved by the short characteristic method \citep{OK87}.  
The outgoing intensity is calculated separately in a similar way for a narrow region and with high resolution by using the pre-calculated level population and thermal structure.

To clarify the importance of the Doppler shifts in the calculation of models and spectra, a few test models that cover our parameter range have been calculated for three cases.
Case 1: 
The Doppler shifts are taken into account only for the outgoing spectrum, while the thermal structure and level populations are calculated without the shifts. The set of angles $\mu$ in this case are defined by the three-point Gaussian quadrature rule (as in \citealt{Dodin15}). 
Case 2:
The Doppler shifts are included in the calculations of the level populations and outgoing spectrum; the thermal structure is taken from Case 1.
Case 3: 
The shifts are considered at all stages.
The comparison between Cases 1 and 2 shows that the outgoing fluxes $H_\nu$ of helium lines can differ by up to a factor of 2, while the same for Cases 2 and 3 gives a difference less than 10 per cent. Therefore, we can use with sufficient accuracy Case 2.

\subsection{Non-equilibrium effects}

We suppose that the both regions are in radiative equilibrium, and the level populations of elements treated in non-LTE are in statistical equilibrium. As we will see later (see Figs. \ref{ions}, \ref{levels}, and \ref{hc}), these conditions are satisfied for the hotspot, because on its upper boundary the cooling gas reaches these equilibriums with sufficient accuracy. In the case of the free-falling gas this question deserves a separate consideration. We suppose that the pre-shock gas is initially cold and neutral, therefore some time is needed to heat and ionize it.
To study the delay in the heating and ionization of the pre-shock gas, the temperature is calculated by using the modified $\Lambda$-correction scheme \citep{Dodin17}, and the level populations are calculated by solving Eq. (\ref{se}). It has been shown by a few test models that the shift of the temperature distribution is significant only for models with very low accretion flux $F_{\rm acc}\approx\rho_0V_0^3/2$, but even in this case the line profiles of H and He do not differ from those calculated in models in radiative equilibrium. The non-equilibrium effects in the level populations appear only in the \ion{H}{i} region, in which helium lines are not formed. Note that the size of the \ion{H}{ii} region remains practically the same as in the equilibrium case. 
Therefore we can conclude that the non-equilibrium effects are more noticeable for low accretion fluxes $F_{\rm acc}$, but for our parameter range they are small and will be ignored.

\subsection{Model limitations}

As in \citet{Dodin15} we use a one-dimensional plane-parallel approximation, therefore we can consider only those layers for which the thickness $H$ is smaller than the stellar radius $R\sim10^{11}$\,cm and the accretion column width $D\sim 10^{10}-10^{11}$\,cm, because the larger the ratio $H/D$, the more radiation escapes through the sides of the accretion column. Although we will consider models with $H = 10^{10} - 4\times10^{10}$\,cm, all quantitative conclusions will concern the lines that are formed in layers with $H\lesssim10^9$\,cm, except for extreme models of the lowest density $N_0 = 10^{10}$\,cm$^{-3}$ and the highest velocity $V_0=500$\,km\,s$^{-1}$.

Generally speaking, there is a velocity gradient in the free-falling gas such that the velocity difference between the upper and lower boundaries of the slab is $0.5V_0H/R.$   In the case of the helium line formation region with $H\sim10^9$\,cm (see Fig. \ref{models}), the difference is about 2\,km\,s$^{-1},$ which is comparable with the thermal velocities in the gas, therefore we suppose that we can neglect the velocity gradient.

We consider the steady-state problem, and therefore cannot directly take into account non-stationary effects of the shock on the spectrum of the hotspot or the pre-shock gas.

\section{Atomic data}
The atomic data used for the calculations of the pre-shock gas and hotspot remain the same as in \citet{Dodin15}, except for the electron collisional excitation rates of \ion{He}{ii} for which more accurate data from \citet{Aggarwal} are used.

The following atoms and ions are included in the calculations of the cooling flow:
\ion{H}{i--ii}, 
\ion{He}{i--iii}, 
\ion{C}{ii--vii}, 
\ion{N}{ii--viii}, 
\ion{O}{ii--ix}, 
\ion{Ne}{ii--xi}, 
\ion{Mg}{ii--xii}, 
\ion{Si}{ii--xiv}, 
\ion{S}{ii--xvi}, 
\ion{Fe}{iii--xix}. 
The atomic data for hydrogen and helium are identical to the data used in the calculation of the hotspot. 
The remaining atoms and ions are treated in a simplified manner, i.e. the most important low levels are considered explicitly, while the upper levels are combined into superlevels in such a way that the total number of levels for each ion does not exceed 20. Energy levels and radiative atomic data are taken from the NORAD\footnote{\url{http://norad.astronomy.ohio-state.edu}} and the TOPbase\footnote{\url{http://cdsweb.u-strasbg.fr/topbase/topbase.html}}. About 3\,000 spectral lines are taken into account in the calculations of the radiative cooling and spectra. Because the photoionization cross-sections include autoionizing resonances, the detailed balance relation gives the total (radiative plus dielectronic) recombination rates for each level. For some ions the main contribution to the total dielectronic recombination is due to the high lying levels, however in the dense plasma of the cooling flow, these levels either do not exist or the recombinations on them are negligible against the high rates of collisional ionization and three-body recombination. Therefore, the total recombination rate to all included levels of an ion must be less than the total recombination rate for the same ion in the interstellar medium. 

Collisional excitation and ionization rates are taken from CHIANTI~v.7.1 \citep{Dere,Landi} for all available transitions. The rest of the rates are calculated by using approximate formulae from \citet{Seaton, Allen} and \citet{vanReg}.

The optimal number of atomic levels for each ion (except ions of hydrogen and helium) is still an open problem.
In this regard, we note that significant changes in atomic data used in the calculations of the cooling flow (e.g. different number of atomic levels; all collisional rates are calculated using the approximate formulae or more precise tables) do not lead to noticeable changes in the spectra of the pre-shock gas or the hotspot, while the particular lines in the ionizing spectra can differ by up to a factor of a few. So we suppose that our atomic models are suitable for the calculation of the ionizing radiation, which can be used to model spectra of the pre-shock gas and the hotspot. However, to study the individual spectral lines forming in the cooling flow, separate detailed research of the corresponding atomic models is needed.

\section{Results}

\subsection{Structure and spectrum of the post-shock region}

The structure of the cooling flow is calculated for $N_0 = 10^{10}-10^{14}$\,cm$^{-3},$ $V_0 = 150-500$\,km\,s$^{-1},$ and the solar elemental abundances \citep{GS98}. 

\begin{figure}
\begin{center}
\includegraphics[scale=0.5]{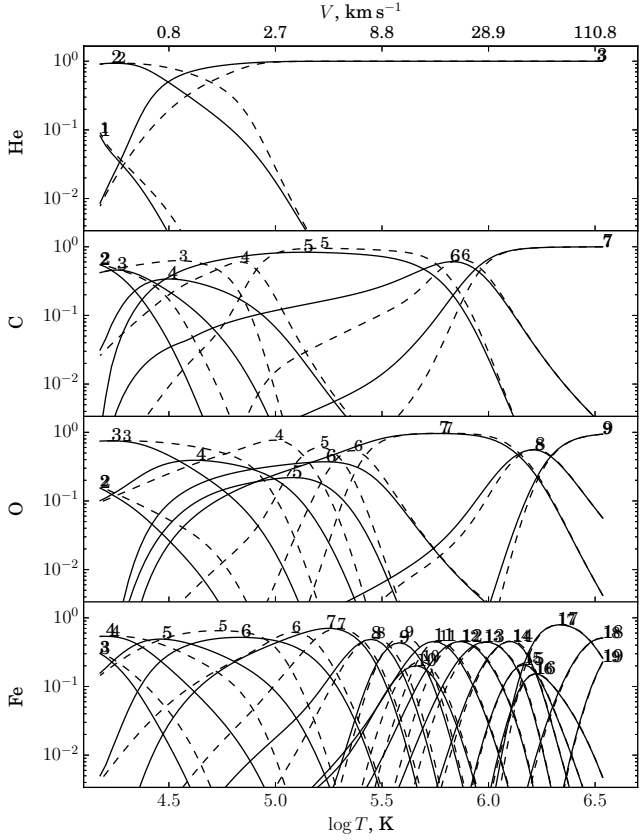}
\end{center}
\caption{The ion fractions $f_i$ of some elements in the cooling flow for $V_0=500$\,km\,s$^{-1}$, $N_0=10^{12}$\,cm$^{-3}$. The dashed curves and reduced font correspond to the model, calculated assuming statistical equilibrium. The solid lines are for the non-equilibrium level populations. The numbers denote the ionization stages, e.g. the number 17 on the lower panel corresponds to the \ion{Fe}{xvii}.
}\label{ions}
\end{figure}

The ion fractions in the cooling region for some elements are shown on Fig. \ref{ions} for two cases:
one where the model (i.e. the temperature, the level populations, the radiation field, etc.) is calculated taking into account departures from the statistical equilibrium (SE), and one where the model is calculated assuming SE.
Generally speaking, the differences in the ion fractions reflect not only non-equilibrium effects, but also related changes in the radiation field, which is especially important for the cooled gas, where photoionization by the radiation of overlying layers plays a significant role. However, in practice, this effect turns to be small, because in the hot region, where the ionizing radiation is formed, the departures from SE are small (see Fig. \ref{ions}). As can be seen from Fig. \ref{ions}, while the gas cools down, the ion fractions converge to the equilibrium values that allows us to neglect the non-equilibrium effects and consider deeper layers as the hotspot.

\begin{figure}
\begin{center}
\includegraphics[scale=0.5]{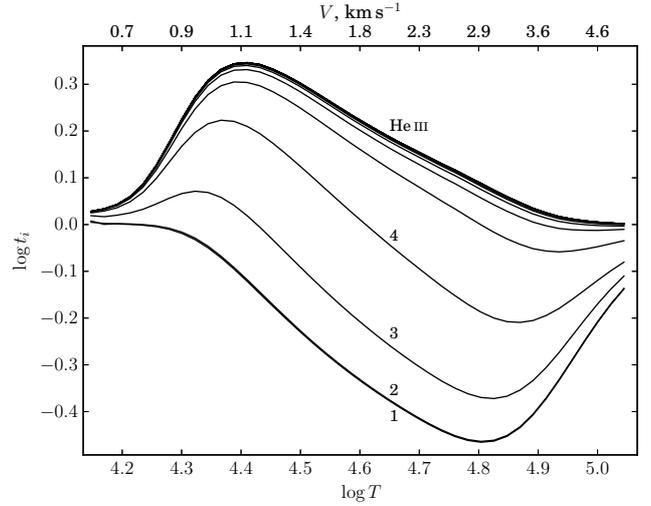}
\end{center}
\caption{The departures of level populations of \ion{He}{ii} from the statistical equilibrium $t_i=n^{\rm neq}_i/n^{\rm eq}_i.$ The structure of the post-shock region is calculated for the non-equilibrium case for $V_0=300$\,km\,s$^{-1}$, $N_0=10^{12}$\,cm$^{-3}.$ The numbers at the curves correspond to the principal quantum number $n$ of the \ion{He}{ii} levels.
The ground state and first excited level have equal departures. The upper line corresponds to the \ion{He}{iii} ion.
The upper axis indicates the gas velocity.}\label{levels}
\end{figure} 

Since in this paper we are primarily concerned with helium spectra, we give a more rigorous and quantitative description of the non-equilibrium effects for each level of \ion{He}{ii} (\ion{He}{i} lines are formed mostly in the hotspot). 
We can simultaneously calculate the non-equilibrium level populations $n^{\rm neq}_i$ and the equilibrium ones $n^{\rm eq}_i$ for a  fixed radiation field and other parameters. The ratios $t_i=n^{\rm neq}_i/n^{\rm eq}_i$, plotted on Fig.~\ref{levels}, show that the departures from the equilibrium at the lower boundary of the cooling region are less than 10 per cent for all levels.

It is remarkable that there is a smooth transition in the t-factors from the ground state of \ion{He}{ii} to \ion{He}{iii}.
Only the first excited level is well coupled to the ground state, while the departures from SE of upper levels correspond to the departures of \ion{He}{iii} due to high rates of collisional ionization and three-body recombination.
The most important levels with $n=3,4$ show intermediate departures from SE. The non-equilibrium effects are usually considered only in calculations of the ion fractions, while the level populations are derived from a balance between collisional excitation from the ground state and radiative decay, i.e. it is assumed that the t-factors of all levels are equal and correspond to the ground state. As can be seen from Fig.~\ref{levels}, in the case of accretion shock this is true only for the first excited level of \ion{He}{ii}.

\begin{figure}
\begin{center}
\includegraphics[scale=0.5]{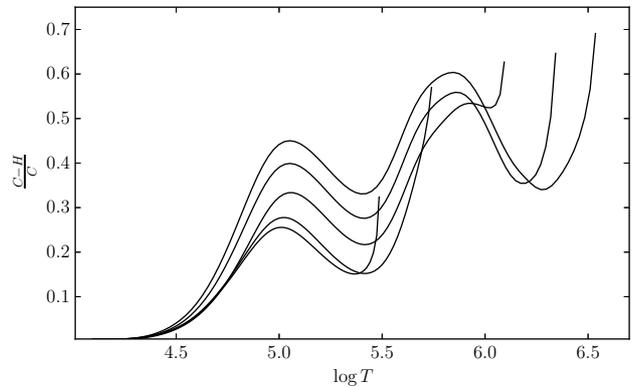}
\end{center}
\caption{The imbalance between heating $H$ and cooling $C$ of the post-shock gas for the models with $N_0=10^{12}$\,cm$^{-3}$,
$V_0=$ 150, 200, 300, 400, 500\,km\,s$^{-1}$. }\label{hc}
\end{figure}

The effects of self-irradiation of the cooling flow lead not only to the photoionization of lower layers by upper ones but also to heating. The ratio between the cooling and heating in the cooling flow is shown in Fig.~\ref{hc}.
It can be seen that the balance between the heating and cooling is reached at the lower boundary of the cooling flow, therefore, deeper layers can be considered in the radiative equilibrium, i.e. as the hotspot. The non-monotonicity of the curves is associated with the appearance and disappearance of strong opacity sources. Note that even in the upper layers there is significant heating.

Accounting for the non-equilibrium effects and self-irradiation leads to changes in the cooling curve.
We define the net cooling rate as $C-H$ that is the rate of change of total energy $E$ per unit mass due to the radiation (see Eqs. \ref{en}--\ref{endef}), while the change of $\varepsilon$ is usually included in the cooling rate, therefore, for comparison purposes, we define a quantity
\begin{equation}\label{muk}
\Lambda = \frac{2.11\times10^{-24} \, \text{g}}{N_e} \left( C-H+u_0\frac{d\varepsilon}{dm} \right).
\end{equation}
Note that in the considered problem the contribution of the last term turns out to be small.

\begin{figure}
\begin{center}
\includegraphics[scale=0.5]{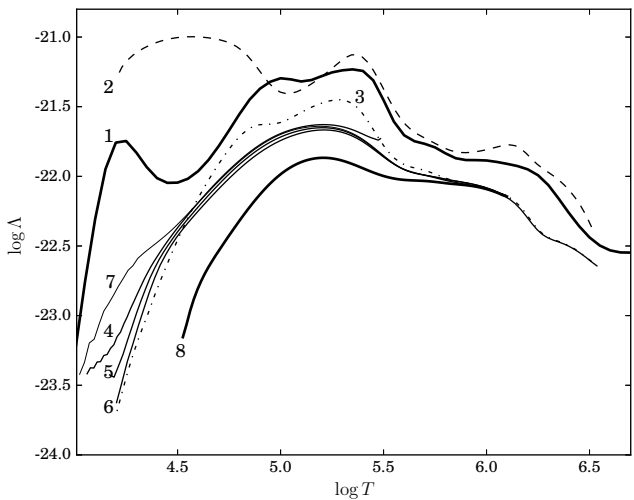}
\end{center}
\caption{The cooling curves $\Lambda(T)$ (erg\,s$^{-1}$\,cm$^{3}$).
Curve (1) is the standard cooling curve of astrophysical plasma in the collisional ionization equilibrium calculated using CHIANTI (the low-density limit, an optically thin gas, no photoionization);
(2) is the curve calculated for $V_0=$500\,km\,s$^{-1},$ $N_0=10^{12}$\,cm$^{-3}$ assuming the statistical equilibrium, 
the heating is ignored $H=0$, effects of photoionization and radiative transfer are taken into account;
(3) is the same curve, but the heating is included; 
(4, 5, 6) are the non-equilibrium curves for $V_0=$150, 300, 500\,km\,s$^{-1},$ $N_0=10^{12}$\,cm$^{-3}$;
(7, 8) are the non-equilibrium curves for $V_0=300$\,km\,s$^{-1},$  $N_0=10^{10},$ $10^{14}$\,cm$^{-3}$ }\label{cooling}
\end{figure} 

The standard cooling curve for a low-density plasma and a few curves for the post-shock region are plotted on Fig.~\ref{cooling}. Curve (2) calculated for the equilibrium level populations and without heating ($H=0$) coincides with the standard one (Curve 1) for $T>10^5$\,K. At lower temperatures there are differences, which are caused by the following effects.

We calculate total heating and cooling, which are integrated over all frequencies, including optically thick frequencies, where  cooling is well balanced by heating in each grid cell. Therefore, by zeroing the heating, we overestimate the cooling rate that leads to the appearance a hump in Curve (2) at $\log T =4.5$ in comparison with Curve (1). 
If these optically thick frequencies are removed from the cooling, then the obtained curve turns out to be similar to the standard one, but the carbon peak at $\log T = 5.0$ is shifted to $\log T =4.8,$ which is caused by photoionization, shifting the ionization equilibrium towards the lower temperatures.

The role of the heating term is illustrated by Curve (3). All the other curves (4-8) correspond to the post-shock region and calculated, accounting for the photoionization, heating, and non-equilibrium effects.
The shape of the cooling curve in the non-equilibrium case is generally consistent with the results obtained by \citet{SD93}. Increasing the gas density (Curves 7-5-8) increases the heating that reduces the net cooling in the lower layers.
Variations of the pre-shock velocity  (Curves 4-5-6) practically do not change the cooling curve, which allows the use of the curve calculated for $V_0=500$\,km\,s$^{-1}$ as an universal curve instead of the standard one in the time-dependent hydrodynamic modelling of accretion impacts.

The negative slope of the standard cooling curve leads to the shock instability (see Introduction); note in connection with this that at moderate densities, the shape of the cooling curve at $\log T >5.5$ does not differ significantly from the standard one. Thus, the heating and non-equilibrium effects do not solve the problem of absence of observational evidences of the shock oscillations.

The simple analytical estimate of the vertical scale $L_{post}$ of the post-shock with radiative cooling gives $L_{post}\propto\rho_0^{-1}$ (see e.g. \citealt{Costa17}). In principle, an increasing heating with increasing density can distort this relation; however, it turned out that the cooling curve depends on density only at low temperatures $\log T <5.5,$ while the cooling length $L_{post}$ is mostly determined by the high-temperature part of $\Lambda(T).$ This results in the vertical size of the cooling region in our models corresponding to the relation $L_{post}\propto\rho_0^{-1}$ with an accuracy better than 10 per cent.

\subsection{Structure and spectrum of the hotspot and the pre-shock gas}\label{results}

In order to calculate the structure and spectrum of the hotspot and the in-falling gas, in addition to the parameters of the shock ($N_0$, $V_0$, the solar elemental abundances) we should define the atmospheric parameters: effective temperature $T_{\rm eff} = 4\,000$\,K, surface gravity $\log g = 4.0,$ and microturbulence velocity $\xi_t = 2.0$\,km\,s$^{-1}.$ The post-shock gas radiates equally in both directions except some optically thick lines, for which the flux directed  towards the star is smaller than the flux directed towards the in-falling gas. Despite the fact that the densities in the pre-shock gas and hotspot differ by almost three orders of magnitude, the temperatures in both regions turn out to be close.

\begin{figure}
\begin{center}
\includegraphics[scale=0.50]{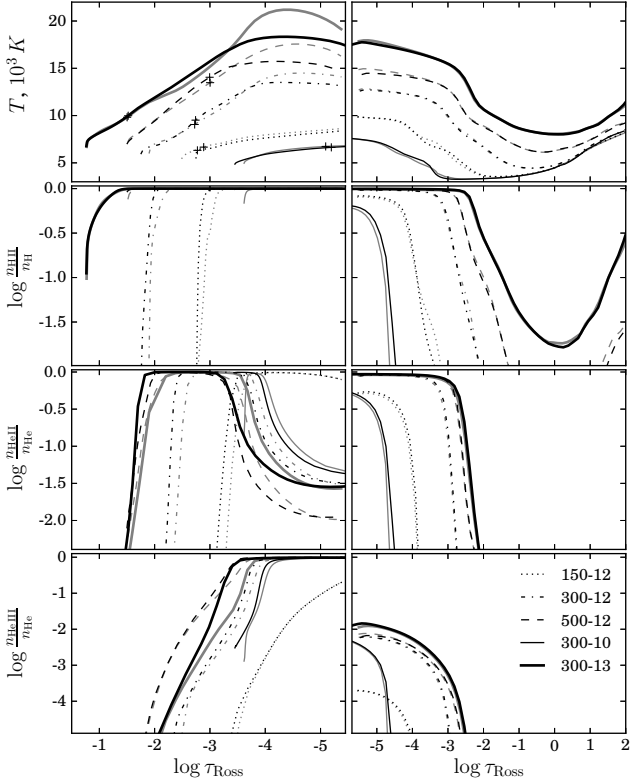}
\end{center}
\caption{
The temperature profiles and ion fractions for hydrogen and helium in the pre-shock region (left panels) and in the hotspot (right panels). The abscissa is the Rosseland-mean optical depth, measured from the shock front for the left panels and the lower boundary of the post-shock region for the right panels. The grey lines are for the HHe model; the black lines are for the HHeCO model.
Different line styles are for different accretion parameters, reported in the in the legend by using the notation $V_0-\log N_0$, where $V_0$ and $N_0$ are measured in km\,s$^{-1}$ and cm$^{-3},$ respectively. The crosses mark the distance of  $10^9$\,cm from the shock front.}\label{models}
\end{figure}

\begin{figure}
\begin{center}
\includegraphics[scale=0.50]{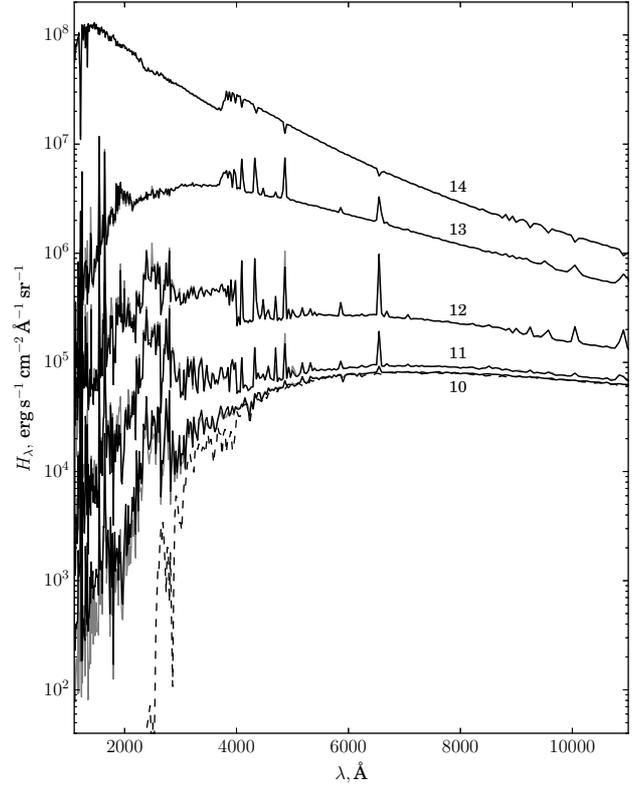}
\end{center}
\caption{ The solid lines are low-resolution ($R\approx150$) spectra of the accretion region at various $\log N_0$ (indicated by numbers at the corresponding lines) and $V_0=300$\,km\,s$^{-1}$. The grey lines are for the HHe model; the black lines are for the HHeCO model. The grey and black lines almost coincide with each other, except for small differences in the UV. The dashed line is for the spectrum of the underlying star with $T_{\rm eff}=4\,000$\,K and $\log g =4.0.$  }\label{sed}
\end{figure}

Fig. \ref{models} shows the temperature profiles and the ion fractions for hydrogen and helium. The typical density profiles for the hotspot can be found in \citet{Dodin15}, and the density in the pre-shock region is constant. The formation region of \ion{He}{i}, \ion{He}{ii} lines for most models is less than $10^9$\,cm (except for extreme models of the lowest density $N_0 = 10^{10}$\,cm$^{-3}$ and the highest velocity $V_0=500$\,km\,s$^{-1}$), which satisfies our assumptions.
As can be seen in Fig.~\ref{sed}, the pre-shock gas is heated up to $10\,000 - 20\,000$\,K, which agrees with earlier calculations \citep{CG98, Lamzin98}. The models, shown on the figure by the grey lines, are calculated taking into the account departures from LTE for hydrogen and helium, while the other elements are treated in LTE. We will label this model as HHe.
The main non-LTE effect is overionization of ions by the hot radiation from the post-shock, which leads to the appearance of \lq{hot ions}\rq{} like \ion{He}{iii}, which do not correspond to the temperatures, reached in the pre-shock or hotspot. The overionization is reduced with increasing density due to three-body recombination putting the populations into LTE.
Similar effects must appear not only for H and He, but also for other elements. This can affect the distribution of the physical parameters in the models ($T$, $N_e$, the radiation field $J_{\nu},$ etc.).
To estimate changes, induced by non-LTE effects in metals, we include the non-LTE treatment of the ions of C and O, using the compilation of atomic data from \citet{Dodin15}. We will label this model as HHeCO. These models turn out to be cooler by $300-3\,000$\,K in the hottest regions near the shock, but slightly hotter ($<600$\,K) in more distant regions. The more important difference is the larger sizes of the hydrogen and helium ionization zones. The comparison of emergent line profiles shows that these changes do not impact the profiles of \ion{H}{i} or \ion{He}{i} but can enhance the red-shifted component of \ion{He}{ii} lines (see. \ref{reshe2}).
A more detailed study of non-LTE effects in metals will be presented in a subsequent work, since in addition to the indirect effect on the \ion{He}{ii} lines, the metal lines are also of independent interest for the explanation of UV and optical spectra.


\begin{landscape}
\begin{figure}
\centerline{\epsfig{file=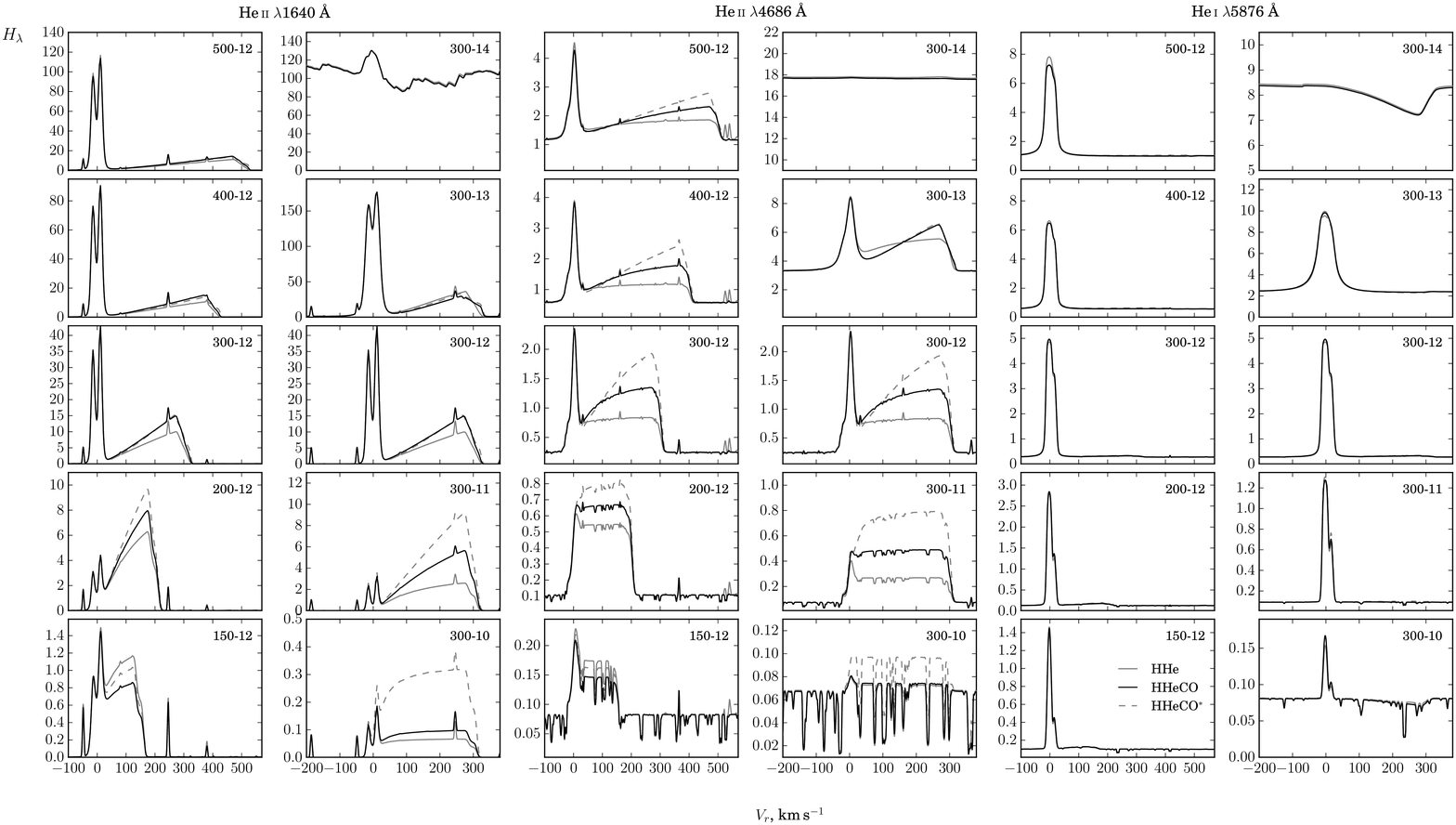, scale=0.5}} 
\caption{Profiles of some significant lines of \ion{He}{i} and \ion{He}{ii} for various accretion parameters.
The vertical axis is the Eddington flux $H_{\lambda}$ in $10^{6}$\,erg\,s$^{-1}$\,sm$^{-2}$\,{\AA}$^{-1}$\,sr$^{-1};$
the horizontal axis is the radial velocity in km\,s$^{-1}.$
The left panels are for  $N_0=10^{12}$\,cm$^{-3},$ $V_0=$150, 200, 300, 400, 500\,km\,s$^{-1}$ (see the $V_0-\log N_0${} keys in the upper-right corner). The right panels are for $V_0=300$\,km\,s$^{-1},$ $N_0=$ $10^{10}$, $10^{11}$, $10^{12}$, $10^{13}$, $10^{14}$\,cm$^{-3}$.
The grey and black solid lines are for HHe and HHeCO models, respectively. The dashed lines are for HHeCO$^*$ models.
}
\label{profiles}
\end{figure}
\end{landscape}

When the models of the pre-shock region, cooling gas, and hotspot are calculated, they are combined into one and the final outgoing spectrum $I_{\lambda}(\mu)$ is calculated. For presentation purposes, we integrate these intensities to obtain the Eddington fluxes $H_{\lambda}=0.5\int_{-1}^1{I_{\lambda}(\mu)\mu d \mu},$ which are shown on Fig.~\ref{sed} and  for the most important lines, on Fig.~\ref{profiles}. Recall that the total flux of a star with the accretion spot can be calculated only when the spot geometry, velocity field, and line of sight are defined. In particular, the shape of the red-shifted component is very sensitive to these parameters.

\subsubsection{ Hydrogen spectrum.}
The behaviour of the central component of hydrogen lines coincides with that described in \citet{Dodin15}. In the present work, we take into account the in-falling gas, but in the case of hydrogen lines we do not fully cover the region of their formation, hence the red-shifted component for some models may be underestimated. We will now describe qualitatively the behaviour of the H~$\beta$ line. A weak red-shifted emission is seen at low densities of the in-falling gas. Somewhere between the densities $N_0$ of $10^{12}$\,cm$^{-3}$  and $10^{13}$\,cm$^{-3}$ the emission is replaced by absorption, which is superimposed on the extremely broad central component. At densities $N_0\sim10^{14}$\,cm$^{-3}$, the central component is completely absorbed in the in-falling gas, in which a deep red-shifted component is formed. Due to the high density, this component is significantly broadened by the Stark effect (${\rm FWHM} \approx600$\,km\,s$^{-1}$). Until now, such broad absorption wings of hydrogen lines have not been observed, which apparently indicates the absence of a very dense gas ($\sim 10^{14}$\,cm$^{-3}$) in the accretion columns of CTTS.

The behaviour of the Balmer jump is analogous to the behaviour of hydrogen lines. The emission Balmer jump corresponds to the low-density models, and the absorption Balmer jump is seen at extremely high densities $N_0\sim10^{14}$\,cm$^{-3}.$ The transition from one type to another occurs at $N_0\sim 10^{13}$\,cm$^{-3},$ when the jump does not emerge. This complements results obtained by Calvet and Gullbring (see fig. 3 of \citealt{Ingleby13}), where only emission jumps arise.

\subsubsection{ He\,{\sevensize I} lines.}
The results for the narrow component of \ion{He}{i} lines are similar to the results of previous work \citep{Dodin15}.
One of the most interesting is that the centre of the \ion{He}{i} $\lambda$5876\,{\AA} line depends on the accretion parameters.
This is due to the fact that the line consists of several fine-structure components, a sequential saturation of which 
shifts its average position. In Fig.\,\ref{profiles} the central wavelength of the lines is taken as the $gf$-weighted average wavelength of all fine-structure components. In addition to this effect, the component can be shifted due to blending with the red-shifted component.

The narrow component of the \ion{He}{i} $\lambda$5876\,{\AA} line is frequently used to measure the magnetic field in the accretion spots (see, for example, \citealt*{Donati, Dodin13}). The measured quantity is the longitudinal field weighted with the line intensity. Thus, the angular dependence of the intensity of the \ion{He}{i} line is of interest for the interpretation of the obtained results, in particular for (magnetic) Doppler imaging. Our models show that the \ion{He}{i} $\lambda$5876\,{\AA} line is optically thick, and the intensity $I(\mu)$ in the line centre is almost isotropic.

The \ion{He}{i} $\lambda$5876\,{\AA} line does not show any significant red-shifted absorption at densities $N_0$ less than $10^{13}$\,cm$^{-3}$.   The absorption appears at higher densities, however; in this case the central component of hydrogen lines reaches a FWHM$>1\,000$\,km\,s$^{-1}.$ The presence of such a broad (but shallow) emission component of hydrogen lines in spectra of CTTS with the red-shifted absorption in lines of \ion{He}{i} has not yet been confirmed by observations.

As expected, the in-falling gas turns out to be opaque in the Balmer and Paschen continuum at densities $N_0\gtrsim 10^{13}$\,cm$^{-3}$. The optical thickness of the pre-shock gas in the model with $V_0=300$\,km\,s$^{-1},$ $N_0=10^{14}$\,cm$^{-3}$ is $\tau_{\rm Ross}>1$ and strongly depends on the chosen geometrical thickness of the slab. The spectrum of the in-falling gas can be described as a spectrum of a star with an effective temperature $T_{\rm hs}=(0.5\rho_0V_0^3/\sigma+T_{\rm eff}^4)^{0.25},$ which is shifted to longer wavelengths.

The \ion{He}{i} $\lambda$10830\,{\AA} line shows similar behaviour with the \ion{He}{i} $\lambda$5876\,{\AA} line, but the red-shifted component is more pronounced. The red-shifted absorption, a depth of which relative to the continuum is less than 20 per cent, is present at densities 
$10^{10}-10^{11}$\,cm$^{-3}$ and is replaced by a weak emission at densities $10^{12}-10^{13}$\,cm$^{-3}$.

\begin{figure}
\begin{center}
\includegraphics[scale=0.5]{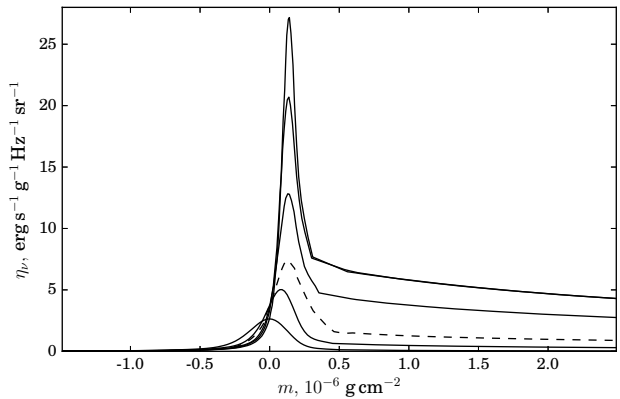}
\end{center}
\caption{The emission coefficient $\eta_{\nu}$ at the line centre of \ion{He}{ii} $\lambda$4686\,{\AA}. The coordinate origin is located at the point where 
the settling velocity is equal to 2\,km\,s$^{-1}.$ The shock front is on the left. The solid lines are for the models with
$V_0=$ 150, 200, 300, 400, 500\,km\,s$^{-1},$ $N_0=10^{12}$\,cm$^{-3}$; the dashed line is for the emission coefficient reduced by a factor of 10 for the model with  $V_0=300$\,km\,s$^{-1},$ $N_0=10^{13}$\,cm$^{-3}$. 
For all models the line is optically thin. It can be seen that the narrow component of the \ion{He}{ii} $\lambda$4686\,{\AA} line is formed at the velocities less than 2\,km\,s$^{-1}$ (i.e. in the right half of the plot). The distribution of $\eta_{\nu}$ in the case of the \ion{He}{ii} $\lambda$1640\,{\AA} line looks similar.}\label{eta4686}
\end{figure} 

\subsubsection{He\,{\sevensize II} lines}\label{reshe2}

There is observational evidence that the narrow component of the \ion{He}{ii} lines is red-shifted by $\sim10$\,km\,s$^{-1}$
(see, for example, \citealt*{Beristain}, \citealt{Petrov, Ardila}). It is believed that the presence of this shift indicates that the component is formed in the cooling flow. However, our models show that such shifts are impossible in the accepted picture of accretion, because at the settling velocities of $\sim10$\,km\,s$^{-1}$ the helium is entirely ionized to \ion{He}{iii} (see Fig.~\ref{ions}). To show where the narrow component is formed in our models, we plot on Fig.~\ref{eta4686} the emission coefficient $\eta_{\nu}$ at the line centre of \ion{He}{ii} $\lambda$4686\,{\AA} near the conventional boundary between the cooling flow and the hotspot. Note firstly a smooth transition of $\eta_{\nu}$ between the two regions. The component turns out to be optically thin for all models, therefore the area under the curve is proportional to the flux, and we can conclude that the component is formed at the settling velocities less than 2\,km\,s$^{-1}.$ If the observed shifts of $\sim10$\,km\,s$^{-1}$ are not related to some simple causes, such as blending with other lines or with the red-shifted component, then they may indicate unaccounted effects like the shock instability, two-dimensional effects \citep{Orlando10, Orlando13} or the flow inhomogeneity \citep{Matsakos,Colombo}.

\ion{He}{ii} lines show a significant red-shifted emission. The observed ratios between the red-shifted and central emissions of \ion{He}{ii} $\lambda$1640\,{\AA} line (fig. 17 of \citealt{Ardila}) agree well with our models for a high velocity and moderate density of $\sim10^{12}$\,cm$^{-3}$. Recall that the profiles shown in Fig. \ref{profiles} correspond to the Eddington flux, in other words, the lines will have such profiles if the whole stellar surface is covered by the accretion spot. The line profiles for different shapes and orientations of the accretion region are different. 

\begin{figure*}
\begin{center}
\includegraphics[scale=0.5]{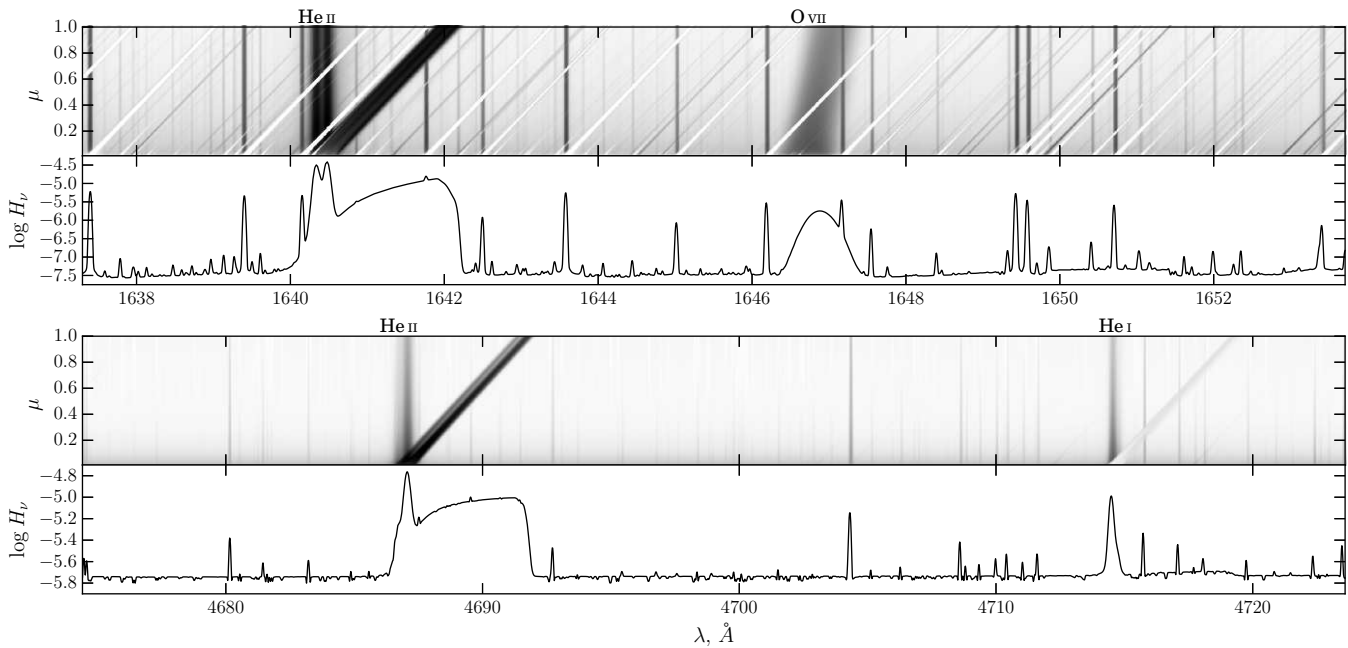}
\end{center}
\caption{The upper parts of the panels are examples of $I_{\nu}(\mu)$ for the UV and visible spectral ranges for the HHeCO model with $V_0=300$\,km\,s$^{-1},$ $N_0=10^{12}$\,cm$^{-3}.$ The inclined strips correspond to the lines that form in the in-falling gas. The flux from the accretion spot is the weighted sum of individual rows, with weights determined by the geometry of the spot. As an example, the Eddington flux (which is proportional to the sum with weights $\mu$) is shown in the bottom parts of the panels. It can be seen that the UV continuum flux is affected by a large number of emission and absorption lines that overlap with each other due to the Doppler effect.}\label{imu}
\end{figure*}

From Fig. \ref{profiles} we can conclude that if the accretion flow contains only a low-density gas, then the red-shifted component dominates in the profile. It is necessary to note that there is some uncertainty in the intensity of this component due to unaccounted non-LTE effects in metals.
The formation region of \ion{He}{ii} lines depends on how deep the ionizing radiation can penetrate, which depends on the ionization of metals as well as on the ionization of helium itself. The overionization of metals can reduce the opacity for the radiation, which is able to ionize \ion{He}{ii} that leads to extending the \ion{He}{iii} zone. Our calculations show that accounting for the non-LTE effects for C and O ions modifies only the red-shifted component of \ion{He}{ii} lines and only at densities $N_0\lesssim10^{12}$\,cm$^{-3}$ (see Fig. \ref{profiles}). 
In order to estimate how the overionization of the other elements can enhance these changes, we calculate the level populations, neglecting in the pre-shock region the opacity due to the elements treated in LTE, which leads to the maximum possible size of the ionization zones of helium ions. The corresponding spectra are shown in Fig. \ref{profiles} and labelled as HHeCO$^*.$

Let us finally discuss the line-blanketing effect. Significant Doppler shifts related to the in-falling gas lead to strong blending of numerous spectral lines, especially in the UV (see Fig. \ref{imu}). The blanketing effect grows with increasing density of the in-falling gas and at $N_0>10^{13}$\,cm$^{-3}$  cannot be ignored even in the visible range.

\section{Conclusions}

In this paper, we have adapted the well-developed methods of the theory of stellar atmospheres to model the accretion shock and its surroundings. We have found the following:
 
(i) The non-equilibrium effects and the self-irradiation modify the shape of the cooling curve only for temperatures  $T<3\times 10^{5}$\,K. Thus, the unstable part of the cooling curve is not changed. Due to small non-equilibrium effects at high temperatures, the cooling curve does not depend significantly on the accretion parameters ($N_0,$ $V_0$), which allows it to be used in numerical simulations of the instability in the same way as the standard curve.

(ii) The departures of the populations of excited levels of \ion{He}{ii} from the statistical equilibrium for a dense plasma of the cooling flow do not correspond to the departures of the ground state but take intermediate values between the departures of the ground state of \ion{He}{ii} and \ion{He}{iii}.

(iii) The central as well as the red-shifted component of hydrogen lines is broadened by the Stark effect, which cannot be neglected in gas kinematic studies. 

(iv) The narrow component of the most important helium lines is optically thick, which leads to a shift of the central wavelength of multicomponent lines (e.g. the \ion{He}{i} $\lambda$5876\,{\AA} line).

(v) An unexpected result is the absence of a visible red-shifted absorption in the \ion{He}{i} $\lambda$5876\,{\AA} line at acceptable accretion parameters, while such an absorption is seen in 4 of 22 CTTS \citep{Beristain}.  
It is probable that this absorption is formed in the accretion column outside the region heated by the shock radiation. However, in this case, we need another ionizing source sufficient for the excitation of helium lines in the accretion column. The probable candidate for the role of the ionizing source may be the coronal X-ray emission \citep{KR11} or the radiation arising at the magnetospheric boundary \citep{Castro}. 

(vi) The model of the steady-state accretion shock predicts that the narrow component of \ion{He}{ii} line has almost zero radial velocity, while the measurements show a redshift of a few km\,s$^{-1}.$ 
\citet{Colombo} have suggested a model of fragmented accretion streams and calculated the profiles of \ion{C}{iv}, which show significant Doppler shifts. It is possible that a similar mechanism is responsible for the observed shifts of the narrow component of \ion{He}{ii}. However, it is necessary to note that \citet{Colombo} have used the approximation of the collisional ionization equilibrium (CIE), while we have seen that the non-equilibrium effects and photoionization shift the formation region of the \ion{C}{iv} lines towards the lower temperatures. Namely the peak ion fraction of \ion{C}{iv} in the CIE case is achieved at $10^5$\,K (the CHIANTI calculations), while accounting for the above-mentioned effects shifts it to $3\times10^4$\,K (see Fig.~\ref{ions}).

(vii) The observed two-component profiles of \ion{He}{ii} $\lambda$1640\,{\AA} line agree well with the modelled ones calculated for a high velocity $V_0=400-500$\,km\,s$^{-1}$ and moderate density of $N_0\sim10^{12}$\,cm$^{-3}.$ 
It motivates performing similar calculations for other elements (primarily C, N, O), lines of which show red-shifted features, because the strong dependence of the shape of such profiles on the flow geometry opens new possibilities for the diagnostics  of the accretion zone.

The cooling curves and intensities $I_{\nu}(\mu)$ for hydrogen and helium lines are publicly available via the project website \url{http://lnfm1.sai.msu.ru/~davnv/hotspot}.

\section*{Acknowledgements}
The author thanks an anonymous referee for valuable comments and suggestions,  which have improved this manuscript.
The reported study was funded by RFBR according to the research project  16-32-00794 mol\_a.

\bibliographystyle{mnras}
\bibliography{dodin2017}

\begin{thebibliography}{}
\makeatletter
\relax
\def\mn@urlcharsother{\let\do\@makeother \do\$\do\&\do\#\do\^\do\_\do\%\do\~}
\def\mn@doi{\begingroup\mn@urlcharsother \@ifnextchar [ {\mn@doi@}
  {\mn@doi@[]}}
\def\mn@doi@[#1]#2{\def\@tempa{#1}\ifx\@tempa\@empty \href
  {http://dx.doi.org/#2} {doi:#2}\else \href {http://dx.doi.org/#2} {#1}\fi
  \endgroup}
\def\mn@eprint#1#2{\mn@eprint@#1:#2::\@nil}
\def\mn@eprint@arXiv#1{\href {http://arxiv.org/abs/#1} {{\tt arXiv:#1}}}
\def\mn@eprint@dblp#1{\href {http://dblp.uni-trier.de/rec/bibtex/#1.xml}
  {dblp:#1}}
\def\mn@eprint@#1:#2:#3:#4\@nil{\def\@tempa {#1}\def\@tempb {#2}\def\@tempc
  {#3}\ifx \@tempc \@empty \let \@tempc \@tempb \let \@tempb \@tempa \fi \ifx
  \@tempb \@empty \def\@tempb {arXiv}\fi \@ifundefined
  {mn@eprint@\@tempb}{\@tempb:\@tempc}{\expandafter \expandafter \csname
  mn@eprint@\@tempb\endcsname \expandafter{\@tempc}}}

\bibitem[\protect\citeauthoryear{{Aggarwal}, {Igarashi}, {Keenan}  \&
  {Nakazaki}}{{Aggarwal} et~al.}{2017}]{Aggarwal}
{Aggarwal} K.,  {Igarashi} A.,  {Keenan} F.,   {Nakazaki} S.,  2017, \mn@doi
  [Atoms] {10.3390/atoms5020019}, \href
  {http://adsabs.harvard.edu/abs/2017Atoms...5...19A} {5, 19}

\bibitem[\protect\citeauthoryear{{Alencar} et~al.,}{{Alencar}
  et~al.}{2012}]{Alencar12}
{Alencar} S.~H.~P.,  et~al., 2012, \mn@doi [\aap]
  {10.1051/0004-6361/201118395}, \href
  {http://adsabs.harvard.edu/abs/2012A%26A...541A.116A} {541, A116}

\bibitem[\protect\citeauthoryear{{Allen}}{{Allen}}{1973}]{Allen}
{Allen} C.~W.,  1973, {Astrophysical quantities}

\bibitem[\protect\citeauthoryear{{Ardila} et~al.,}{{Ardila}
  et~al.}{2013}]{Ardila}
{Ardila} D.~R.,  et~al., 2013, \mn@doi [\apjs] {10.1088/0067-0049/207/1/1},
  \href {http://adsabs.harvard.edu/abs/2013ApJS..207....1A} {207, 1}

\bibitem[\protect\citeauthoryear{{Beristain}, {Edwards}  \& {Kwan}}{{Beristain}
  et~al.}{2001}]{Beristain}
{Beristain} G.,  {Edwards} S.,   {Kwan} J.,  2001, \mn@doi [\apj]
  {10.1086/320233}, \href {http://adsabs.harvard.edu/abs/2001ApJ...551.1037B}
  {551, 1037}

\bibitem[\protect\citeauthoryear{{Blinova}, {Romanova}  \&
  {Lovelace}}{{Blinova} et~al.}{2016}]{Blinova16}
{Blinova} A.~A.,  {Romanova} M.~M.,   {Lovelace} R.~V.~E.,  2016, \mn@doi
  [\mnras] {10.1093/mnras/stw786}, \href
  {http://adsabs.harvard.edu/abs/2016MNRAS.459.2354B} {459, 2354}

\bibitem[\protect\citeauthoryear{Brown, Byrne  \& Hindmarsh}{Brown
  et~al.}{1989}]{Brown}
Brown P.~N.,  Byrne G.~D.,   Hindmarsh A.~C.,  1989, SIAM J. Sci. Stat.
  Comput., 10, 1038

\bibitem[\protect\citeauthoryear{{Calvet} \& {Gullbring}}{{Calvet} \&
  {Gullbring}}{1998}]{CG98}
{Calvet} N.,  {Gullbring} E.,  1998, \mn@doi [\apj] {10.1086/306527}, \href
  {http://adsabs.harvard.edu/abs/1998ApJ...509..802C} {509, 802}

\bibitem[\protect\citeauthoryear{{Colombo}, {Orlando}, {Peres}, {Argiroffi}  \&
  {Reale}}{{Colombo} et~al.}{2016}]{Colombo}
{Colombo} S.,  {Orlando} S.,  {Peres} G.,  {Argiroffi} C.,   {Reale} F.,  2016,
  \mn@doi [\aap] {10.1051/0004-6361/201628858}, \href
  {http://adsabs.harvard.edu/abs/2016A%26A...594A..93C} {594, A93}

\bibitem[\protect\citeauthoryear{{Costa}, {Orlando}, {Peres}, {Argiroffi}  \&
  {Bonito}}{{Costa} et~al.}{2017}]{Costa17}
{Costa} G.,  {Orlando} S.,  {Peres} G.,  {Argiroffi} C.,   {Bonito} R.,  2017,
  \mn@doi [\aap] {10.1051/0004-6361/201628554}, \href
  {http://adsabs.harvard.edu/abs/2017A%26A...597A...1C} {597, A1}

\bibitem[\protect\citeauthoryear{{Dere}, {Landi}, {Mason}, {Monsignori Fossi}
  \& {Young}}{{Dere} et~al.}{1997}]{Dere}
{Dere} K.~P.,  {Landi} E.,  {Mason} H.~E.,  {Monsignori Fossi} B.~C.,   {Young}
  P.~R.,  1997, \mn@doi [\aaps] {10.1051/aas:1997368}, \href
  {http://adsabs.harvard.edu/abs/1997A%26AS..125..149D} {125}

\bibitem[\protect\citeauthoryear{{Dodin}}{{Dodin}}{2015}]{Dodin15}
{Dodin} A.~V.,  2015, \mn@doi [Astronomy Letters] {10.1134/S1063773715050023},
  \href {http://adsabs.harvard.edu/abs/2015AstL...41..196D} {41, 196}

\bibitem[\protect\citeauthoryear{{Dodin}}{{Dodin}}{2017}]{Dodin17}
{Dodin} A.,  2017, in {Balega} Y.~Y.,  {Kudryavtsev} D.~O.,  {Romanyuk} I.~I.,
   {Yakunin} I.~A.,  eds,  Astronomical Society of the Pacific Conference
  Series Vol. 510, Stars: From Collapse to Collapse. p.~110

\bibitem[\protect\citeauthoryear{{Dodin} \& {Lamzin}}{{Dodin} \&
  {Lamzin}}{2012}]{Dodin12}
{Dodin} A.~V.,  {Lamzin} S.~A.,  2012, \mn@doi [Astronomy Letters]
  {10.1134/S1063773712100027}, \href
  {http://adsabs.harvard.edu/abs/2012AstL...38..649D} {38, 649}

\bibitem[\protect\citeauthoryear{{Dodin}, {Lamzin}  \& {Chuntonov}}{{Dodin}
  et~al.}{2013}]{Dodin13}
{Dodin} A.~V.,  {Lamzin} S.~A.,   {Chuntonov} G.~A.,  2013, \mn@doi
  [Astrophysical Bulletin] {10.1134/S1990341313020053}, \href
  {http://adsabs.harvard.edu/abs/2013AstBu..68..177D} {68, 177}

\bibitem[\protect\citeauthoryear{{Donati} et~al.,}{{Donati}
  et~al.}{2007}]{Donati07}
{Donati} J.-F.,  et~al., 2007, \mn@doi [\mnras]
  {10.1111/j.1365-2966.2007.12194.x}, \href
  {http://adsabs.harvard.edu/abs/2007MNRAS.380.1297D} {380, 1297}

\bibitem[\protect\citeauthoryear{{Donati} et~al.,}{{Donati}
  et~al.}{2008}]{Donati08}
{Donati} J.-F.,  et~al., 2008, \mn@doi [\mnras]
  {10.1111/j.1365-2966.2008.13111.x}, \href
  {http://adsabs.harvard.edu/abs/2008MNRAS.386.1234D} {386, 1234}

\bibitem[\protect\citeauthoryear{{Donati} et~al.,}{{Donati}
  et~al.}{2011}]{Donati11}
{Donati} J.-F.,  et~al., 2011, \mn@doi [\mnras]
  {10.1111/j.1365-2966.2010.18069.x}, \href
  {http://adsabs.harvard.edu/abs/2011MNRAS.412.2454D} {412, 2454}

\bibitem[\protect\citeauthoryear{{Donati} et~al.,}{{Donati}
  et~al.}{2013}]{Donati}
{Donati} J.-F.,  et~al., 2013, \mn@doi [\mnras] {10.1093/mnras/stt1622}, \href
  {http://adsabs.harvard.edu/abs/2013MNRAS.436..881D} {436, 881}

\bibitem[\protect\citeauthoryear{{Drake}, {Ratzlaff}, {Laming}  \&
  {Raymond}}{{Drake} et~al.}{2009}]{Drake}
{Drake} J.~J.,  {Ratzlaff} P.~W.,  {Laming} J.~M.,   {Raymond} J.,  2009,
  \mn@doi [\apj] {10.1088/0004-637X/703/2/1224}, \href
  {http://adsabs.harvard.edu/abs/2009ApJ...703.1224D} {703, 1224}

\bibitem[\protect\citeauthoryear{{G{\'o}mez de Castro} \& {von
  Rekowski}}{{G{\'o}mez de Castro} \& {von Rekowski}}{2011}]{Castro}
{G{\'o}mez de Castro} A.~I.,  {von Rekowski} B.,  2011, \mn@doi [\mnras]
  {10.1111/j.1365-2966.2010.17726.x}, \href
  {http://adsabs.harvard.edu/abs/2011MNRAS.411..849G} {411, 849}

\bibitem[\protect\citeauthoryear{{Grevesse} \& {Sauval}}{{Grevesse} \&
  {Sauval}}{1998}]{GS98}
{Grevesse} N.,  {Sauval} A.~J.,  1998, \mn@doi [\ssr]
  {10.1023/A:1005161325181}, \href
  {http://adsabs.harvard.edu/abs/1998SSRv...85..161G} {85, 161}

\bibitem[\protect\citeauthoryear{{G{\"u}nther} et~al.,}{{G{\"u}nther}
  et~al.}{2010}]{Gunther}
{G{\"u}nther} H.~M.,  et~al., 2010, \mn@doi [\aap]
  {10.1051/0004-6361/201013996}, \href
  {http://adsabs.harvard.edu/abs/2010A%26A...518A..54G} {518, A54}

\bibitem[\protect\citeauthoryear{{Hartmann}, {Herczeg}  \& {Calvet}}{{Hartmann}
  et~al.}{2016}]{Hartmann16}
{Hartmann} L.,  {Herczeg} G.,   {Calvet} N.,  2016, \mn@doi [\araa]
  {10.1146/annurev-astro-081915-023347}, \href
  {http://adsabs.harvard.edu/abs/2016ARA%26A..54..135H} {54, 135}

\bibitem[\protect\citeauthoryear{{Ingleby} et~al.,}{{Ingleby}
  et~al.}{2013}]{Ingleby13}
{Ingleby} L.,  et~al., 2013, \mn@doi [\apj] {10.1088/0004-637X/767/2/112},
  \href {http://adsabs.harvard.edu/abs/2013ApJ...767..112I} {767, 112}

\bibitem[\protect\citeauthoryear{{Koenigl}}{{Koenigl}}{1991}]{Koenigl91}
{Koenigl} A.,  1991, \mn@doi [\apjl] {10.1086/185972}, \href
  {http://adsabs.harvard.edu/abs/1991ApJ...370L..39K} {370, L39}

\bibitem[\protect\citeauthoryear{{Koldoba}, {Ustyugova}, {Romanova}  \&
  {Lovelace}}{{Koldoba} et~al.}{2008}]{Koldoba}
{Koldoba} A.~V.,  {Ustyugova} G.~V.,  {Romanova} M.~M.,   {Lovelace} R.~V.~E.,
  2008, \mn@doi [\mnras] {10.1111/j.1365-2966.2008.13394.x}, \href
  {http://adsabs.harvard.edu/abs/2008MNRAS.388..357K} {388, 357}

\bibitem[\protect\citeauthoryear{{Kurosawa} \& {Romanova}}{{Kurosawa} \&
  {Romanova}}{2013}]{KR13}
{Kurosawa} R.,  {Romanova} M.~M.,  2013, \mn@doi [\mnras]
  {10.1093/mnras/stt365}, \href
  {http://adsabs.harvard.edu/abs/2013MNRAS.431.2673K} {431, 2673}

\bibitem[\protect\citeauthoryear{{Kurosawa}, {Romanova}  \&
  {Harries}}{{Kurosawa} et~al.}{2011}]{KR11}
{Kurosawa} R.,  {Romanova} M.~M.,   {Harries} T.~J.,  2011, \mn@doi [\mnras]
  {10.1111/j.1365-2966.2011.19216.x}, \href
  {http://adsabs.harvard.edu/abs/2011MNRAS.416.2623K} {416, 2623}

\bibitem[\protect\citeauthoryear{{Lamzin}}{{Lamzin}}{1995}]{Lamzin95}
{Lamzin} S.~A.,  1995, \aap, \href
  {http://adsabs.harvard.edu/abs/1995A%26A...295L..20L} {295, L20}

\bibitem[\protect\citeauthoryear{{Lamzin}}{{Lamzin}}{1998}]{Lamzin98}
{Lamzin} S.~A.,  1998, Astronomy Reports, \href
  {http://adsabs.harvard.edu/abs/1998ARep...42..322L} {42, 322}

\bibitem[\protect\citeauthoryear{{Lamzin}}{{Lamzin}}{2003}]{Lamzin03}
{Lamzin} S.~A.,  2003, \mn@doi [Astronomy Reports] {10.1134/1.1583777}, \href
  {http://adsabs.harvard.edu/abs/2003ARep...47..498L} {47, 498}

\bibitem[\protect\citeauthoryear{{Landi}, {Young}, {Dere}, {Del Zanna}  \&
  {Mason}}{{Landi} et~al.}{2013}]{Landi}
{Landi} E.,  {Young} P.~R.,  {Dere} K.~P.,  {Del Zanna} G.,   {Mason} H.~E.,
  2013, \mn@doi [\apj] {10.1088/0004-637X/763/2/86}, \href
  {http://adsabs.harvard.edu/abs/2013ApJ...763...86L} {763, 86}

\bibitem[\protect\citeauthoryear{{Matsakos} et~al.,}{{Matsakos}
  et~al.}{2013}]{Matsakos}
{Matsakos} T.,  et~al., 2013, \mn@doi [\aap] {10.1051/0004-6361/201321820},
  \href {http://adsabs.harvard.edu/abs/2013A%26A...557A..69M} {557, A69}

\bibitem[\protect\citeauthoryear{{Muzerolle}, {Calvet}  \&
  {Hartmann}}{{Muzerolle} et~al.}{2001}]{Muzerolle01}
{Muzerolle} J.,  {Calvet} N.,   {Hartmann} L.,  2001, \mn@doi [\apj]
  {10.1086/319779}, \href {http://adsabs.harvard.edu/abs/2001ApJ...550..944M}
  {550, 944}

\bibitem[\protect\citeauthoryear{{Olson} \& {Kunasz}}{{Olson} \&
  {Kunasz}}{1987}]{OK87}
{Olson} G.~L.,  {Kunasz} P.~B.,  1987, \mn@doi [\jqsrt]
  {10.1016/0022-4073(87)90027-6}, \href
  {http://adsabs.harvard.edu/abs/1987JQSRT..38..325O} {38, 325}

\bibitem[\protect\citeauthoryear{{Orlando}, {Sacco}, {Argiroffi}, {Reale},
  {Peres}  \& {Maggio}}{{Orlando} et~al.}{2010}]{Orlando10}
{Orlando} S.,  {Sacco} G.~G.,  {Argiroffi} C.,  {Reale} F.,  {Peres} G.,
  {Maggio} A.,  2010, \mn@doi [\aap] {10.1051/0004-6361/200913565}, \href
  {http://adsabs.harvard.edu/abs/2010A%26A...510A..71O} {510, A71}

\bibitem[\protect\citeauthoryear{{Orlando} et~al.,}{{Orlando}
  et~al.}{2013}]{Orlando13}
{Orlando} S.,  et~al., 2013, \mn@doi [\aap] {10.1051/0004-6361/201322076},
  \href {http://adsabs.harvard.edu/abs/2013A%26A...559A.127O} {559, A127}

\bibitem[\protect\citeauthoryear{{Petrov}, {Gahm}, {Gameiro}, {Duemmler},
  {Ilyin}, {Laakkonen}, {Lago}  \& {Tuominen}}{{Petrov} et~al.}{2001}]{Petrov}
{Petrov} P.~P.,  {Gahm} G.~F.,  {Gameiro} J.~F.,  {Duemmler} R.,  {Ilyin}
  I.~V.,  {Laakkonen} T.,  {Lago} M.~T.~V.~T.,   {Tuominen} I.,  2001, \mn@doi
  [\aap] {10.1051/0004-6361:20010203}, \href
  {http://adsabs.harvard.edu/abs/2001A%26A...369..993P} {369, 993}

\bibitem[\protect\citeauthoryear{{Piskunov}, {Tuominen}  \& {Vilhu}}{{Piskunov}
  et~al.}{1990}]{Piskunov90}
{Piskunov} N.~E.,  {Tuominen} I.,   {Vilhu} O.,  1990, \aap, \href
  {http://adsabs.harvard.edu/abs/1990A%26A...230..363P} {230, 363}

\bibitem[\protect\citeauthoryear{{Romanova}, {Ustyugova}, {Koldoba}, {Wick}  \&
  {Lovelace}}{{Romanova} et~al.}{2003}]{Romanova03}
{Romanova} M.~M.,  {Ustyugova} G.~V.,  {Koldoba} A.~V.,  {Wick} J.~V.,
  {Lovelace} R.~V.~E.,  2003, \mn@doi [\apj] {10.1086/377514}, \href
  {http://adsabs.harvard.edu/abs/2003ApJ...595.1009R} {595, 1009}

\bibitem[\protect\citeauthoryear{{Sacco}, {Argiroffi}, {Orlando}, {Maggio},
  {Peres}  \& {Reale}}{{Sacco} et~al.}{2008}]{Sacco}
{Sacco} G.~G.,  {Argiroffi} C.,  {Orlando} S.,  {Maggio} A.,  {Peres} G.,
  {Reale} F.,  2008, \mn@doi [\aap] {10.1051/0004-6361:200810753}, \href
  {http://adsabs.harvard.edu/abs/2008A%26A...491L..17S} {491, L17}

\bibitem[\protect\citeauthoryear{{Seaton}}{{Seaton}}{1962}]{Seaton}
{Seaton} M.,  1962, in 'Atomic and Molecular Processes'{}, New York Academic
  Press

\bibitem[\protect\citeauthoryear{{Sutherland} \& {Dopita}}{{Sutherland} \&
  {Dopita}}{1993}]{SD93}
{Sutherland} R.~S.,  {Dopita} M.~A.,  1993, \mn@doi [\apjs] {10.1086/191823},
  \href {http://adsabs.harvard.edu/abs/1993ApJS...88..253S} {88, 253}

\bibitem[\protect\citeauthoryear{{van Regemorter}}{{van
  Regemorter}}{1962}]{vanReg}
{van Regemorter} H.,  1962, \mn@doi [\apj] {10.1086/147445}, \href
  {http://adsabs.harvard.edu/abs/1962ApJ...136..906V} {136, 906}

\makeatother
\end{thebibliography}
\appendix

\section{The jump conditions}\label{A1}

To obtain the gas parameters in the first point behind the shock front, we rewrite the conservation equations (\ref{mass}--\ref{endef}) as: 
\begin{equation}\label{appmass}
       \rho_{\rm in} V_{\rm in} = \rho_0 V_0,   
\end{equation}       
%
\begin{equation}\label{appmom}
       P_{\rm in}+\rho_{\rm in} V_{\rm in}^2 = \rho_0 V_0^2 + P_0, 
\end{equation} 
%
\begin{equation}\label{appen}
     \frac{V_{\rm in}^2}{2} + \frac{5}{2}\frac{P_{\rm in}}{\rho_{\rm in}}+\varepsilon_{\rm in} = \frac{V_0^2}{2} + \frac{5}{2}\frac{P_0}{\rho_0}+\varepsilon_0. 
\end{equation}     
The indices \lq0\rq{} and  \lq{in}\rq{} denote the last point of the pre-shock and the first point of the hotspot. We assume that between these states the gas does not lose energy via the radiation, but becomes ionized, i.e. the part of its thermal energy converted to $\varepsilon.$
Expressing $\rho_{\rm in}$  from Eq. \ref{appmass}, $P_{\rm in}$  from Eq. \ref{appmom} and substituting them in Eq. \ref{appen},
we obtain:
\begin{equation}\label{app1}
     \frac{V_{\rm in}^2}{2} + \frac{5}{2}(V_0-V_{\rm in})V_{\rm in}+\frac{5}{2}\frac{P_0}{\rho_0}\frac{V_{\rm in}}{V_0} +\varepsilon_{\rm in} = \frac{V_0^2}{2} + \frac{5}{2}\frac{P_0}{\rho_0}+\varepsilon_0. 
\end{equation}  
If we introduce the variables
\begin{equation}\label{appdef}
x=\frac{V_{\rm in}}{V_0}, \qquad A = \frac{P_0}{\rho_0V_0^2},\qquad  B=\frac{\varepsilon_{\rm in}-\varepsilon_0}{V_0^2},
\end{equation}  
then Eq. \ref{app1} is transformed to the form:
\begin{equation}\label{app2}
x^2-\frac{5}{4}\left(1+A\right)x+\frac{1}{4}+\frac{5}{4}A-\frac{1}{2}B=0.
\end{equation} 
Note that in our case $A,B\ll1,$ then, solving the quadratic equation and retaining only first order terms, we obtain:
\begin{equation}\label{app2}
x_1 \approx 1+\frac{2}{3}B, \qquad x_2 \approx \frac{1}{4}\left(1+5A-\frac{8}{3}B \right).
\end{equation} 
The shock wave corresponds to the second root. The parameters $\rho_{\rm in}$ and $P_{\rm in}$ can be obtained by substituting the initial velocity $V_{\rm in}=x_2V_0$ in Eq.~\ref{appmass} and Eq.~\ref{appmom}.

\bsp	
\label{lastpage}
\end{document}